\RequirePackage{amssymb}
\documentclass[sigconf,natbib=true]{acmart}

\AtBeginDocument{%
  \providecommand\BibTeX{{%
    \normalfont B\kern-0.5em{\scshape i\kern-0.25em b}\kern-0.8em\TeX}}}

\usepackage{setspace}
\usepackage[version=4]{mhchem}
\usepackage{tabularx}
\usepackage[most]{tcolorbox}
\usepackage{bbding}
\usepackage{url}
\setcopyright{acmlicensed}
\copyrightyear{2024}
\acmYear{2024}
\acmDOI{XXXXXXX.XXXXXXX}

\acmConference[SIGIR 2024]{The 47th International ACM SIGIR Conference on Research and Development in Information Retrieval}{July 14--18, 2024}{Washington D.C., USA}
\begin{document}

\title{Reducing the climate impact of data portals: a case study}



\author{Noah Gießing}
\email{Noah.Giessing@fiz-karlsruhe.de}
\orcid{0009-0006-5268-2519}
\affiliation{
  \institution{FIZ Karlsruhe Leibniz Institute for Information Infrastructure}
  \streetaddress{Franklinstrasse 11}
  \city{Berlin}
  \country{Germany}
  \postcode{10587}
}

\author{Madhurima Deb}
\email{Madhurima.Deb@fiz-karlsruhe.de}
\affiliation{
  \institution{FIZ Karlsruhe Leibniz Institute for Information Infrastructure}
  \streetaddress{Franklinstrasse 11}
  \city{Berlin}
  \country{Germany}
  \postcode{10587}
}

\author{Ankit Satpute}
\email{Ankit.Satpute@fiz-karlsruhe.de}
\orcid{0000-0003-3219-026X}
\affiliation{
  \institution{FIZ Karlsruhe Leibniz Institute for Information Infrastructure}
  \streetaddress{Franklinstrasse 11}
  \city{Berlin}
  \country{Germany}
  \postcode{10587}
}


\author{Moritz Schubotz}
\email{Moritz.Schubotz@fiz-karlsruhe.de}
\orcid{0000-0001-7141-4997}
\affiliation{
  \institution{FIZ Karlsruhe Leibniz Institute for Information Infrastructure}
  \streetaddress{Franklinstrasse 11}
  \city{Berlin}
  \country{Germany}
  \postcode{10587}
}

\author{Olaf Teschke}
\email{Olaf.Teschke@fiz-karlsruhe.de}
\orcid{0009-0003-4089-9647}
\affiliation{
  \institution{FIZ Karlsruhe Leibniz Institute for Information Infrastructure}
  \streetaddress{Franklinstrasse 11}
  \city{Berlin}
  \country{Germany}
  \postcode{10587}
}



\renewcommand{\shortauthors}{Gießing, et al.}

\begin{abstract}
The carbon footprint share of the information and communication technology (ICT) sector has steadily increased in the past decade and is predicted to make up as much as 23 \% of global emissions in 2030.
This shows a pressing need for developers, including the information retrieval community, to make their code more energy-efficient. In this project proposal, we discuss techniques to reduce the energy footprint of the MaRDI (Mathematical Research Data Initiative) Portal, a MediaWiki-based knowledge base. In future work, we plan to implement these changes and provide concrete measurements on the gain in energy efficiency. Researchers developing similar knowledge bases can adapt our measures to reduce their environmental footprint. In this way, we are working on mitigating the climate impact of Information Retrieval research.
\end{abstract}




\maketitle

\section{Introduction}
The increasing societal adoption and sophistication of information technology have led to a disproportionate increase in greenhouse gas emissions within the information technology \& communication (ICT) sector. 
This trend is expected to continue in the coming years, as one of the studies~\cite{challe2015} projected a share of between 7 and 23 \% of the global greenhouse emissions in 2030. The growth in emissions is driven in part by Information Retrieval, particularly, especially nowadays, when traditional retrieval models get augmented or replaced by more energy-intensive chat-based solutions such as Bing Chat or Google Gemini.\cite{devries2023growing}
The impact of ICT remains nevertheless under-addressed in public discussions surrounding climate change mitigation as shifting electricity production to renewable energy is expected to drive emissions in the ICT sector down. 
However, such an argument fails to address the emissions produced in the current period, where electricity is still provided to a large extent by carbon-based energy sources. 
Even though enterprises may choose to power their services with green energy only, higher energy consumption will mean that non-green energy is needed to provide electricity for other consumers.
Pledges by companies like Alphabet to offset their emissions through reforestation likewise have been exposed as largely intransparent and ineffective in combating climate change, see for example\cite{probst2023systematic}.
Hence, measures to improve energy efficiency are urgently needed.\footnote{This also covers the case of covering energy use by using an in-house green power plant. While commendable, any saved power can generally be transferred to the national power grid. Therefore, such private investments can still be understood as effectively similar to choosing a green energy provider.} The information retrieval community shares a responsibility to lead the way in reducing their climate impact.

\subsection{Green coding}
The principles of green coding have been well explored in recent studies, such as \cite{junger2024potentials}. 
Design and implementation of carbon-aware web interfaces, real-time indication of energy consumption, minimization of data transfer and usage of energy-intensive features, deduplication and compression of data, and implementation of appropriate distributed design patterns are the essential steps for achieving sustainable computing according to \cite{pazienza2024holistic}. 
Along with that, \cite{lannelongue2021ten} suggests calculating the carbon footprint of the ongoing task, including carbon footprint in the cost-benefit analysis, choosing the right computational facility, and improving the efficiency of the code for greener computational practice. 

\subsection{Project Outline}
This work aims to discuss the existing methodologies and measures to broaden the path of greener and sustainable development of MaRDI, a mathematics-specific knowledge base.
The obtained results could be used as a model to follow for researchers involved in similar portals such as Wikipedia, Open Reviews and other wikis in improving their climate footprint.
In forthcoming work, we will implement these measures and assess their practical reduction in energy demand. 
We will measure the amount of data saved and infer from the reduction in power usage the reduction in carbon footprint. 
The contribution of the proposed project lies in the collective implementation of existing efficiency-increasing techniques and the assessment of the summary impact on efficiency.

\section{MaRDI Portal}
The MaRDI (Mathematical Research Data Initiative)~\footnote{\url{https://portal.mardi4nfdi.de/wiki/Portal}} portal is part of Germany's National Research Data Initiative (NFDI).
The portal focuses on developing and maintaining a mathematics-specific knowledge graph containing publications and research data items such as datasets, software, and mathematical formulae \cite{schubotz2023bravo}.  
The main purpose of MaRDI is to serve as a one-stop sustainable contact point for the mathematical community as well as researchers from other relevant applied disciplines to retrieve mathematical research data. The MaRDI portal is the unified interface for providing services from diverse domains such as computer algebra, scientific computing, statistics, machine learning and interdisciplinary fields. \\
The MaRDI portal is developed on the MediaWiki software, on which Wikipedia is based. 
MediaWiki is an open-source software capable of handling traffic-heavy websites. 
Rich in features, it is effective in terms of performance, user-friendly configuration, and scalability \cite{wiki:xxx}.   
The MaRDI knowledge graph currently contains nearly 450 million triples (i.e., edges in the knowledge graph). 
The portal runs on a single virtual machine created by OpenStack. 
The main memory size is 128 GB, 32 CPUs, and 1500 GB HDD.

\section{Methodology}

\begin{table*}[ht]
\begin{center}
\begin{tabularx}{\textwidth}{||>{\centering\arraybackslash}X | >{\centering\arraybackslash}X | >{\centering\arraybackslash}X||} 
 \hline
 Technique & Description & Performance Tradeoff? \\ [0.5ex] 
 \hline\hline
 Profilers & Check code components for performance bottlenecks & \XSolidBrush \\ 
 \hline
 Caching & Offload frequently requested data to memory & \XSolidBrush \\
 \hline
 Search engine & Choose more power-efficient search engine  & \Checkmark \\
 \hline
 Container usage & Recreate containers instead of restarting & \Checkmark \\
 \hline
 Content delivery network & Collaborative data hosting to reduce hops & \Checkmark\\ [1ex] 
 \hline
  Minification and concatenation & Reduce web site code size by shortening variable names and concatenating documents & \XSolidBrush \\ 
 \hline
 Compression & Zip source code and media files & \XSolidBrush \\
 \hline
 Database optimization & Avoid expensive database searches through e.g. pre-indexing & \XSolidBrush \\
 \hline
 Data saving & Make knowledge graph smaller by removing unused features & \Checkmark \\
 \hline
 Lazy Loading & Load website dynamically, after user interaction & \XSolidBrush\\ [1ex] 
 \hline
 Extension optimization & Replace external web services by natively supported ones & \XSolidBrush\\ [1ex] 
 \hline
 Server configuration & Change server settings regarding capacity & \Checkmark \\ [1ex] 
 \hline
 Hardware optimization & Use more energy-efficient hardware  & \Checkmark \\ [1ex] 
 \hline

\end{tabularx}
\end{center}
\caption{Overview of techniques for enhancement in energy efficiency}
\end{table*}
In \ref{Solutions}, we present the methods by which we will aim to make the MaRDI portal more energy-efficient. \ref{measurement} then discusses how we can measure the emission reductions, once they are implemented.
\subsection{Probable solutions using green coding}
\label{Solutions}
\begin{center}
\end{center}
The following solutions can be roughly categorized as based on either of the following four principles of green coding:
\begin{enumerate}
    \item Make delivered data smaller.
    \item Do not deliver unnecessary data.
    \item Ensure optimized availability for often-requested data.
    \item Make internal workflows more efficient.
\end{enumerate}

\subsubsection{Profilers}
A common suggestion\cite{junger2024potentials} in green coding is to use profiling tools to identify potential hotspots of energy consumption within software applications. 
These hotspots, once identified, can become focal points for developers to implement algorithmic optimizations aimed at reducing the computational demand and, consequently, the energy footprint of their applications. 
The cited survey paper \cite{junger2024potentials} mentions several such tools, with hardware profilers such as \texttt{nvidia-smi} on one hand and software profilers like PEEK \cite{10.5555/2750315.2750321} on the other.
In addition, such algorithmic optimizations will benefit general performance of MaRDI, so there are no downsides to be considered.
\subsubsection{Caching}
Caching is a strategic approach that involves offloading data anticipated to be requested in the future to either server or client memory, thereby enabling quicker and more efficient delivery of content~\cite{schubotz2022caching}. 
The theoretical basis for caching is the principle of locality of reference, a concept suggesting that users are likely to request previously accessed data (temporal locality) or to make requests confined to specific clusters of data (spatial or branch locality). 
These patterns allow caching systems to optimize the data retrieval processes significantly.
HTTP accelerators, tools designed to optimize the transfer and processing of HTTP requests between clients and servers, utilize these caching principles to enhance performance. 
Such accelerators dramatically reduce the load on web servers and decrease response times by storing copies of files or outputs of dynamic requests in memory. 
Among the HTTP accelerators that support the MediaWiki platform, which powers MaRDI, are Zend OpCache and Varnish. 
In periods of high traffic they manage to mitigate server load by serving cached content directly to users without the server's need for continuous data processing.
The greater efficiency of data delivery also contributes to a broader resource optimization strategy. 
\subsubsection{Search engine}
The current search engine used by MaRDI is Elasticsearch, which is the current industry standard for database search. 
A drawback of Elasticsearch is that it is Java-based, which does not allow for lower-level code optimization.
This Java dependency limits opportunities for lower-level code optimizations that might be accessible in languages closer to the system hardware, like C or C++.
A potential alternative would be Manticore Search\footnote{\url{https://github.com/manticoresoftware/manticoresearch}}, based on the earlier Sphinx search engine and written in C++. 
This efficiency should translate into faster search response times and reduced resource consumption, which are factors for large-scale search operations.
However, it would have to be investigated whether Manticore's search precision meets the necessary standards for search accuracy and relevancy.
This process would involve benchmarking Manticore Search against Elasticsearch using real-world datasets and search scenarios to ensure that the alternative does not compromise the quality of search results while potentially offering greater efficiency.

\subsubsection{Optimized container usage}
MaRDI utilizes Docker containers for managing its computational environment. At runtime, the containers can be interacted with via a web interface through the software Portainer\footnote{\url{https://www.portainer.io/}}. 
This setup offers a modular approach to application deployment, which is both scalable and efficient.
The dissertation by Kreten \cite{Kreten2022} studies various optimization techniques that could significantly enhance the resource efficiency of container usage within environments like MaRDI's. 
Among these techniques, two noteworthy recommendations stand out: firstly, the strategy to cache container images in memory, which helps in reducing the load times and resource demands when containers are frequently created and destroyed. 
Secondly, there is a preference for recreating containers rather than merely restarting them. 
The experiments cited in Kreten's dissertation indicate that implementing these optimization techniques can lead to an energy saving of approximately 12\%. 
\subsubsection{Content delivery network}
As an extension of the caching principle, a peer-to-peer-based content delivery network would load website components from nearby clients who had previously loaded them instead of sending data from a centralized location. 
This way, the number of hops required to access content can be reduced, resulting in reduced energy demand due to hopping.
Tips for decentralizing Wikipedia and other wikis are discussed in \cite{urdaneta2009wikipedia}.

\subsubsection{Minification and concatenation}
Minification and concatenation both aim to reduce the computational load to serve and load content without changing the website functionality. Reduced load means less energy consumption on both the server and client sides.
Minification achieves this aim by reducing the size of code files. 
This process involves several methods, such as removing unnecessary whitespace, comments, and formatting from the code and renaming variables to shorter, less descriptive names. 
These alterations do not change the code's functionality but decrease the file size, resulting in smaller website that are faster to load and less burdensome on network resources.
The practical impacts of minification have been empirically studied by Sakamoto et al. \cite{sakamoto2015empirical}. 
The study's findings suggest that the application of website minification strategies can reduce page sizes ranging from 39\% to 45\%. 
Another effective strategy alongside minification is the concatenation of multiple CSS or JavaScript files into a single, larger file. 
This method is beneficial because it reduces the number of HTTP requests a browser must make to load a webpage.
\subsubsection{Compression}
Another method to minimize the amount of data transmitted over the network is the compression of website data through tools such as GZIP or Brotli.
When a user requests a web page, they receive a compressed version of the content, which their browser then decompresses or ``unpack'' locally. 
This process reduces the size of data transmitted, speeding up the download time, reducing bandwidth usage, and thus lessening the load on web servers.
These algorithms are particularly effective for text-based content such as the source code of web pages. 
As per Sakamoto et al. \cite{sakamoto2015empirical},
compression can reduce web content size by up to 65\%.
\subsubsection{Database optimization}
By tailoring the database design to accommodate frequently appearing request patterns, search operations can be made more efficient, reducing both the processing time and resource consumption associated with data retrieval.
One effective method to achieve this is through pre-indexing certain columns within the database. 
Indexing involves creating a data structure that allows for faster retrieval of records. 
By indexing columns that are known to be often queried, the database system can quickly locate the starting point of the data sought without scanning every record, thus avoiding the need to read through the entire database. 
Additionally, caching popular requests in a rapidly accessible storage layer results can further optimize performance. 
\subsubsection{Data saving techniques}
We will investigate how a more economical graph design can reduce the size of the knowledge graph. Similarly, we check whether there are unnecessary or unused features we can delete to decrease the size of individual websites.

\subsubsection{Lazy Loading}
Lazy Loading refers to only partially loading web pages, pending further interactions. 
Lazy loading is often used in the case of social networks: these websites rely on providing quasi-unlimited content to the user in an uninterrupted way. 
Loading all the contents would induce high latency, so the web page continuously loads new elements when scrolling. 
In the MaRDI knowledge graph, certain pages might contain lots of data, e.g., an author's entire publication list. 
Lazy Loading can save bandwidth since a user might be interested only in the most cited publications.

\subsubsection{Extension optimization}\label{extensions}
Using natively supported extensions is preferred for added efficiency: We can save energy when requesting external web services. Furthermore, natively supported modules are optimized for performance within the given system. 

\subsubsection{Server configuration}
Server settings, particularly load management, are another variable influencing energy consumption. 
Here, we also have to consider temporal variation within the energy mix: during daytime and on more windy days, a much larger share of German electricity stems from renewables, so a possible policy would be to allow more requests during these times. This is studied, in terms of job scheduling by \cite{de2019green}, e.g..

\subsubsection{Hardware optimization}
Selecting the right hardware components, such as storage devices and CPUs, can lead to significant improvements in energy consumption, as well as overall system responsiveness.
Starting with storage, Solid State Drives (SSDs) represent a superior choice over Hard Disk Drives (HDDs) in many aspects, particularly in terms of energy efficiency. 
SSDs, which utilize flash memory to store data, consume only about a tenth of the energy required by HDDs. 
For instance, in the setup of MaRDI, integrating SSDs can significantly reduce the energy footprint of data storage operations.
Besides, the faster data access speeds of SSDs also contribute to reduced latency and quicker data retrieval times.\\  
Arguably, the choice of CPU is a more significant factor influencing power consumption and computational efficiency, as processor energy consumption is usually the dominant factor in a server's power consumption. 
Modern CPUs with advanced power management features can adjust their operating frequencies based on the current load, which helps to reduce power usage during periods of low activity. The survey paper \cite{heinisch2020towards} indicates that for specific tasks, the energy consumption of different CPUs might vary by a factor of up to 4. For specific tasks, higher efficiency could also be achieved by offloading to GPUs
\subsection{Efficiency measurement}\label{measurement}
A service for calculating a website's energy footprint is provided by the Beacon website~\footnote{\url{https://digitalbeacon.co/}}. 
Beacon measures the size of various website components and the number of times each element is called. 
From a fixed byte-to-\ce{CO2}-emission rate, the total website emission is extrapolated. 
A more in-depth calculation would involve observing the server load in handling specific requests. 

\section{Conclusion}
In this work, we drew on existing research in efficient and green coding for finding solutions to make our mathematical knowledge graph MaRDI more green. In this way, we address the pressing need for the ICT sector to reduce its energy footprint. 
Many of the suggested methods should also improve service quality by lowering response latency as a side effect, reminding us that economic and ecological aims need not be mutually exclusive. 
In continuing our research, we plan to implement these solutions and produce measurements on the actual reduced energy footprint. We hope our experiences can inform other researchers in information retrieval who aim to reduce their environmental footprint.

\section*{Acknowledgements}
This work was funded by DFG grant numbers 460135501,  437179652, the Deutscher Akademischer Austauschdienst (DAAD, German Academic Exchange Service - 57515245), the Lower Saxony Ministry of Science and Culture and the VW Foundation.

\bibliographystyle{ACM-Reference-Format}
\bibliography{main}


\begin{thebibliography}{15}


\ifx \showCODEN    \undefined \def \showCODEN     #1{\unskip}     \fi
\ifx \showDOI      \undefined \def \showDOI       #1{#1}\fi
\ifx \showISBNx    \undefined \def \showISBNx     #1{\unskip}     \fi
\ifx \showISBNxiii \undefined \def \showISBNxiii  #1{\unskip}     \fi
\ifx \showISSN     \undefined \def \showISSN      #1{\unskip}     \fi
\ifx \showLCCN     \undefined \def \showLCCN      #1{\unskip}     \fi
\ifx \shownote     \undefined \def \shownote      #1{#1}          \fi
\ifx \showarticletitle \undefined \def \showarticletitle #1{#1}   \fi
\ifx \showURL      \undefined \def \showURL       {\relax}        \fi
\providecommand\bibfield[2]{#2}
\providecommand\bibinfo[2]{#2}
\providecommand\natexlab[1]{#1}
\providecommand\showeprint[2][]{arXiv:#2}

\bibitem[Andrae and Edler(2015)]%
        {challe2015}
\bibfield{author}{\bibinfo{person}{Anders S.~G. Andrae} {and}
  \bibinfo{person}{Tomas Edler}.} \bibinfo{year}{2015}\natexlab{}.
\newblock \showarticletitle{On Global Electricity Usage of Communication
  Technology: Trends to 2030}.
\newblock \bibinfo{journal}{\emph{Challenges}} \bibinfo{volume}{6},
  \bibinfo{number}{1} (\bibinfo{year}{2015}), \bibinfo{pages}{117--157}.
\newblock
\showISSN{2078-1547}
\urldef\tempurl%
\url{https://doi.org/10.3390/challe6010117}
\showDOI{\tempurl}


\bibitem[De~Courchelle et~al\mbox{.}(2019)]%
        {de2019green}
\bibfield{author}{\bibinfo{person}{In{\`e}s De~Courchelle},
  \bibinfo{person}{Tom Gu{\'e}rout}, \bibinfo{person}{Georges Da~Costa},
  \bibinfo{person}{Thierry Monteil}, {and} \bibinfo{person}{Yann Labit}.}
  \bibinfo{year}{2019}\natexlab{}.
\newblock \showarticletitle{Green energy efficient scheduling management}.
\newblock \bibinfo{journal}{\emph{Simulation Modelling Practice and Theory}}
  \bibinfo{volume}{93} (\bibinfo{year}{2019}), \bibinfo{pages}{208--232}.
\newblock


\bibitem[de~Vries(2023)]%
        {devries2023growing}
\bibfield{author}{\bibinfo{person}{Alex de Vries}.}
  \bibinfo{year}{2023}\natexlab{}.
\newblock \showarticletitle{The growing energy footprint of artificial
  intelligence}.
\newblock \bibinfo{journal}{\emph{Joule}} \bibinfo{volume}{7},
  \bibinfo{number}{10} (\bibinfo{year}{2023}), \bibinfo{pages}{2191--2194}.
\newblock


\bibitem[Heinisch et~al\mbox{.}(2020)]%
        {heinisch2020towards}
\bibfield{author}{\bibinfo{person}{Philip Heinisch}, \bibinfo{person}{Katharina
  Ostaszewski}, {and} \bibinfo{person}{Hendrik Ranocha}.}
  \bibinfo{year}{2020}\natexlab{}.
\newblock \showarticletitle{Towards green computing: A survey of performance
  and energy efficiency of different platforms using opencl}.
\newblock \bibinfo{journal}{\emph{arXiv preprint arXiv:2003.03794}}
  (\bibinfo{year}{2020}).
\newblock


\bibitem[H\"{o}nig et~al\mbox{.}(2014)]%
        {10.5555/2750315.2750321}
\bibfield{author}{\bibinfo{person}{Timo H\"{o}nig}, \bibinfo{person}{Heiko
  Janker}, \bibinfo{person}{Christopher Eibel}, \bibinfo{person}{Wolfgang
  Schr\"{o}der-Preikschat}, \bibinfo{person}{Oliver Mihelic}, {and}
  \bibinfo{person}{R\"{u}diger Kapitza}.} \bibinfo{year}{2014}\natexlab{}.
\newblock \showarticletitle{Proactive energy-aware programming with PEEK}. In
  \bibinfo{booktitle}{\emph{Proceedings of the 2014 International Conference on
  Timely Results in Operating Systems}} (Broomfield, CO)
  \emph{(\bibinfo{series}{TRIOS'14})}. \bibinfo{publisher}{USENIX Association},
  \bibinfo{address}{USA}, \bibinfo{pages}{6}.
\newblock


\bibitem[Junger et~al\mbox{.}(2024)]%
        {junger2024potentials}
\bibfield{author}{\bibinfo{person}{Dennis Junger}, \bibinfo{person}{Max
  Westing}, \bibinfo{person}{Christopher~P Freitag}, \bibinfo{person}{Achim
  Guldner}, \bibinfo{person}{Konstantin Mittelbach}, \bibinfo{person}{Kira
  Oberg{\"o}ker}, \bibinfo{person}{Sebastian Weber}, \bibinfo{person}{Stefan
  Naumann}, {and} \bibinfo{person}{Volker Wohlgemuth}.}
  \bibinfo{year}{2024}\natexlab{}.
\newblock \showarticletitle{Potentials of Green Coding--Findings and
  Recommendations for Industry, Education and Science--Extended Paper}.
\newblock \bibinfo{journal}{\emph{arXiv preprint arXiv:2402.18227}}
  (\bibinfo{year}{2024}).
\newblock


\bibitem[Kreten(2022)]%
        {Kreten2022}
\bibfield{author}{\bibinfo{person}{Sandro Kreten}.}
  \bibinfo{year}{2022}\natexlab{}.
\newblock \emph{\bibinfo{title}{Modellbildung und Umsetzung von Methoden zur
  energieeffizienten Nutzung von Containertechnologien}}.
\newblock doctoralthesis. \bibinfo{school}{Universit{\"a}t Trier}.
\newblock
\urldef\tempurl%
\url{https://doi.org/10.25353/ubtr-xxxx-099b-6fe5}
\showDOI{\tempurl}


\bibitem[Lannelongue et~al\mbox{.}(2021)]%
        {lannelongue2021ten}
\bibfield{author}{\bibinfo{person}{Lo{\"\i}c Lannelongue},
  \bibinfo{person}{Jason Grealey}, \bibinfo{person}{Alex Bateman}, {and}
  \bibinfo{person}{Michael Inouye}.} \bibinfo{year}{2021}\natexlab{}.
\newblock \bibinfo{title}{Ten simple rules to make your computing more
  environmentally sustainable}.
\newblock , \bibinfo{numpages}{e1009324}~pages.
\newblock


\bibitem[MediaWiki(2024)]%
        {wiki:xxx}
\bibfield{author}{\bibinfo{person}{MediaWiki}.}
  \bibinfo{year}{2024}\natexlab{}.
\newblock \bibinfo{title}{Manual:What is MediaWiki? --- MediaWiki{,}}.
\newblock
\newblock
\urldef\tempurl%
\url{https://www.mediawiki.org/w/index.php?title=Manual:What_is_MediaWiki%3F&oldid=6364594}
\showURL{%
\tempurl}
\newblock
\shownote{[Online; accessed 19-April-2024]}.


\bibitem[Pazienza et~al\mbox{.}(2024)]%
        {pazienza2024holistic}
\bibfield{author}{\bibinfo{person}{Andrea Pazienza}, \bibinfo{person}{Giovanni
  Baselli}, \bibinfo{person}{Daniele~Carlo Vinci}, {and}
  \bibinfo{person}{Maria~Vittoria Trussoni}.} \bibinfo{year}{2024}\natexlab{}.
\newblock \showarticletitle{A holistic approach to environmentally sustainable
  computing}.
\newblock \bibinfo{journal}{\emph{Innovations in Systems and Software
  Engineering}} (\bibinfo{year}{2024}), \bibinfo{pages}{1--25}.
\newblock


\bibitem[Probst et~al\mbox{.}(2023)]%
        {probst2023systematic}
\bibfield{author}{\bibinfo{person}{Benedict Probst}, \bibinfo{person}{Malte
  Toetzke}, \bibinfo{person}{Andreas Kontoleon}, \bibinfo{person}{Laura
  Diaz~Anadon}, {and} \bibinfo{person}{Volker~H Hoffmann}.}
  \bibinfo{year}{2023}\natexlab{}.
\newblock \showarticletitle{Systematic review of the actual emissions
  reductions of carbon offset projects across all major sectors}.
\newblock  (\bibinfo{year}{2023}).
\newblock


\bibitem[Sakamoto et~al\mbox{.}(2015)]%
        {sakamoto2015empirical}
\bibfield{author}{\bibinfo{person}{Yasutaka Sakamoto},
  \bibinfo{person}{Shinsuke Matsumoto}, \bibinfo{person}{Seiki Tokunaga},
  \bibinfo{person}{Sachio Saiki}, {and} \bibinfo{person}{Masahide Nakamura}.}
  \bibinfo{year}{2015}\natexlab{}.
\newblock \showarticletitle{Empirical study on effects of script minification
  and HTTP compression for traffic reduction}. In
  \bibinfo{booktitle}{\emph{2015 Third International Conference on Digital
  Information, Networking, and Wireless Communications (DINWC)}}. IEEE,
  \bibinfo{pages}{127--132}.
\newblock


\bibitem[Schubotz et~al\mbox{.}(2023)]%
        {schubotz2023bravo}
\bibfield{author}{\bibinfo{person}{Moritz Schubotz}, \bibinfo{person}{Eloi
  Ferrer}, \bibinfo{person}{Johannes Stegm{\"u}ller}, \bibinfo{person}{Daniel
  Mietchen}, \bibinfo{person}{Olaf Teschke}, \bibinfo{person}{Larissa Pusch},
  {and} \bibinfo{person}{Tim~OF Conrad}.} \bibinfo{year}{2023}\natexlab{}.
\newblock \showarticletitle{Bravo MaRDI: A Wikibase Powered Knowledge Graph on
  Mathematics}.
\newblock \bibinfo{journal}{\emph{arXiv preprint arXiv:2309.11484}}
  (\bibinfo{year}{2023}), \bibinfo{pages}{1--14}.
\newblock


\bibitem[Schubotz et~al\mbox{.}(2022)]%
        {schubotz2022caching}
\bibfield{author}{\bibinfo{person}{Moritz Schubotz}, \bibinfo{person}{Ankit
  Satpute}, \bibinfo{person}{Andr{\'e} Greiner-Petter}, \bibinfo{person}{Akiko
  Aizawa}, {and} \bibinfo{person}{Bela Gipp}.} \bibinfo{year}{2022}\natexlab{}.
\newblock \showarticletitle{Caching and Reproducibility: Making Data Science
  Experiments Faster and FAIRer}.
\newblock \bibinfo{journal}{\emph{Frontiers in Research Metrics and Analytics}}
   \bibinfo{volume}{7} (\bibinfo{year}{2022}), \bibinfo{pages}{861944}.
\newblock


\bibitem[Urdaneta et~al\mbox{.}(2009)]%
        {urdaneta2009wikipedia}
\bibfield{author}{\bibinfo{person}{Guido Urdaneta}, \bibinfo{person}{Guillaume
  Pierre}, {and} \bibinfo{person}{Maarten Van~Steen}.}
  \bibinfo{year}{2009}\natexlab{}.
\newblock \showarticletitle{Wikipedia workload analysis for decentralized
  hosting}.
\newblock \bibinfo{journal}{\emph{Computer Networks}} \bibinfo{volume}{53},
  \bibinfo{number}{11} (\bibinfo{year}{2009}), \bibinfo{pages}{1830--1845}.
\newblock


\end{thebibliography}
\end{document}